\documentclass[fleqn,10pt]{wlscirep}
\usepackage[utf8]{inputenc}
\usepackage[T1]{fontenc}

\usepackage{graphicx}
\usepackage{dcolumn}
\usepackage{bm}
\usepackage{color}
\usepackage{booktabs}
\usepackage[center]{subfigure}

\usepackage[normalem]{ulem}
\usepackage{soul}

\title{Correlated-photon imaging at 10 volumetric images per second}

\author[1,2,4]{Gianlorenzo Massaro}
\author[3,4]{Paul Mos}
\author[2,4]{Sergii Vasiukov}
\author[2,5]{Francesco Di Lena}
\author[1,2,5]{Francesco Scattarella}
\author[1,2,*]{Francesco V. Pepe}
\author[3]{Arin Ulku}
\author[1,2]{Davide Giannella}
\author[3]{Edoardo Charbon}
\author[3,6]{Claudio Bruschini}
\author[1,2,6]{Milena D'Angelo}

\affil[1]{Universit\`a degli studi di Bari, Dipartimento Interuniversitario di Fisica, Bari, I-70126, Italy}
\affil[2]{Istituto Nazionale di Fisica Nucleare, Sezione di Bari, Bari, I-70125, Bari, Italy}
\affil[3]{Ecole polytechnique f\'ed\'erale de Lausanne (EPFL), Neuch\^atel, 2002, Switzerland}
\affil[4]{Equal contribution}
\affil[5]{Equal contribution}
\affil[6]{Equal contribution}

\affil[*]{francesco.pepe@ba.infn.it}

\begin{abstract}
The correlation properties of light provide an outstanding tool to overcome the limitations of traditional imaging techniques. A relevant case is represented by correlation plenoptic imaging (CPI), a quantum-inspired volumetric imaging protocol employing spatio-temporally correlated photons from either entangled or chaotic sources to address the main limitations of conventional light-field imaging, namely, the poor spatial resolution and the reduced change of perspective for 3D imaging. However, the application potential of high-resolution imaging modalities relying on photon correlations is limited, in practice, by the need to collect a large number of frames. This creates a gap, unacceptable for many relevant tasks, between the time performance of correlated-light imaging and that of traditional imaging methods. In this article, we address this issue by exploiting the photon number correlations intrinsic in \textit{chaotic} light, combined with a cutting-edge ultrafast sensor made of a large array of single-photon avalanche diodes (SPADs). This combination of source and sensor is embedded within a novel single-lens CPI scheme enabling to acquire 10 \textit{volumetric} images per second. Our results place correlated-photon imaging at a competitive edge and prove its potential in practical applications. 
\end{abstract}

\begin{document}

\flushbottom
\maketitle

\thispagestyle{empty}

\section*{Introduction}

The very first studies on the peculiar properties of entangled photons from spontaneous parametric down-conversion (SPDC) \cite{klyshko1988effect,rubin1996transverse} led to the discovery of quite counter-intuitive correlation imaging modalities \cite{klyshko1988effect,pittman1995optical,pittman1996two}, and to the development of so-called quantum imaging. In this context, spatio-temporal correlations of light have been exploited 
both toward novel imaging schemes and to overcome the limits of traditional imaging (e.g., in terms of resolution, spectral range, contrast, signal-to-noise ratio) \cite{genovese2016real,moreau2019imaging,gilabertebasset2019perspectives}. On the one hand, quantum correlated photons have been employed to achieve otherwise unattainable quantum effects, such as imaging with undetected photons \cite{lemos2014quantum} and sub-shot-noise imaging \cite{brida2009measurement,brida2010experimental,samantaray2017realization,meda2017photon}. On the other hand, it was discovered that specific correlation imaging tasks could still be achieved when replacing SPDC photons with \textit{chaotic} light sources, such as thermal and pseudo-thermal \cite{bennink2002two,valencia2005two,gatti2004ghost,ferri2005high,scarcelli2006can}. 
Such classically correlated sources typically entails a worse signal-to-noise ratio (SNR) than expected with entangled photons \cite{osullivan2010comparison,brida2011systematic}, but also brings in several practical advantages: light sources are simpler and more feasible (potentially, any available incoherent source could serve the purpose \cite{karmakar2012ghost,liu2014lensless}), the image acquisition time can be significantly reduced by avoiding the low production rate of entangled photons, and protocols are potentially insensitive to the deleterious effects of decoherence. A wide range of quantum-inspired imaging protocols based on chaotic light correlations have been demonstrated so far, in different application scenarios: imaging of objects hidden to the main sensor \cite{meyers2008ghost,hardy2010ghost}, dual wavelength imaging \cite{chan2009twocolor,duan2019nondegenerate} (analogous to the implementation with entangled photons \cite{karmakar2010twocolor,kalashnikov2016infrared}), detection of objects surrounded by turbulence \cite{meyers2011turbulence,shi2012adaptive,bina2013backscattering}, 3D imaging through computational ghost imaging \cite{sun20133dcomputational}, refocusing and 3D imaging through correlation plenotpic imaging and microscopy \cite{dangelo2016correlation,pepe2017exploring,pepe2017diffraction,dilena2018correlation,dilena2020correlation,scagliola2020correlation,massaro2022lightfield}.

However, the variety of advantages entailed by the spatio-temporal correlation properties of light clashes with the most relevant challenge of correlation imaging methods: reducing the acquisition time to make them effectively competitive with the corresponding state-of-the-art traditional techniques. 
Such a challenge is intrinsic to this approach, since evaluating correlation functions requires sampling a high enough number of statistical realizations. The use of high-resolution CCD and CMOS cameras in correlation imaging setups \cite{bennink2002two,ferri2005high,aspden2013epr,pepe2017diffraction,moreau2019imaging,gilabertebasset2019perspectives,massaro2022lightfield} has enabled to overcome the extremely time-consuming mechanical scanning of the image plane, as performed in pioneering correlation imaging experiments \cite{pittman1995optical,valencia2005two,scarcelli2006can}. There are two time scales that one must consider when dealing with the speed of correlation imaging with a camera \cite{gilabertebasset2019perspectives}. First, if $N_t$ camera frames are required, the \textit{total acquisition time} is equal to $T_{\mathrm{image}}= N_t / R$, with $R$ the frame rate; its inverse $T_{\mathrm{image}}^{-1}$ is the rate at which correlation images can be acquired. The acquisition rate can be reduced by i) maximizing $R$, as in high-speed sensors, and ii) optimizing the trade-off between number of frames and SNR, with the latter expected to depend on $\sqrt{N_t}$ \cite{erkmen2009signal,osullivan2010comparison,lantz2014optimizing,scala2019signal,massaro2022comparative}. The other crucial time scale to be considered is the \textit{gating time}, namely, the effective sensitivity window in which a single frame is acquired: if it is larger than the source coherence time, intensity fluctuations in each frame are partially erased \cite{mandel1995optical}, making it more difficult to reconstruct their correlations, and thus increasing the required $N_t$. An innovative class of detectors, made of an array of single-photon avalanche photodiodes (SPAD) \cite{antolovic2016photon,ulku2019spad,ulku2020spad,madonini2021}, provides one of the most interesting solutions available to take on both temporal issues. SPAD arrays ensure at once fast acquisition rates of up to about $10^5$ fps, sub-$10$~ns gating time, and low noise: SPAD arrays are characterized by the absence of readout noise, whereas dark counts are limited to less than $10$ counts per second per pixel. The reduced noise is essential for high-quality sampling of the light statistics with fewer frames. SPAD arrays have thus been extensively employed for correlation measurements, and yet more rarely for correlated-photon imaging (see Ref. \cite{madonini2021} for a wide review and detailed bibliography). In this context, despite the development of high-resolution SPAD arrays has permitted to largely cut the acquisition time by exploiting multiple pair coincidences within a single frame, the acquisition times attained in entangled-based high-resolution 2D imaging remain larger than one second \cite{lubin2019quantum,defienne2021fullfield}. In fact, even neglecting the data bandwidth limitation of state-of-the-art high-resolution SPAD arrays, which currently prevents large arrays from being used at full speed for extended periods, the long acquisition time needed by entanglement-based correlation imaging is essentially determined by the low production rate of SPDC.

In this work, we shall demonstrate correlated-photon imaging at a rate of 10 \textit{volumetric} images per second, with a field of view of $256\times 256$ pixels. The key for achieving such performances is the use of: 1) SwissSPAD2 \cite{antolovic2016photon,ulku2019spad,ulku2020spad,abbattista2021towards}, a SPAD array operated at a $512\times 256$-pixel resolution, and capable of acquiring up to $10^5$ frames per second; 2) chaotic light illumination, which  enables to avoid the low speed related with the low production rates of parametric down-conversion; 3) the principles of correlation plenoptic imaging (CPI) \cite{dangelo2016correlation,pepe2016correlation,dilena2018correlation,dilena2020correlation,scagliola2020correlation}, a technique that enables to acquire information about both the spatial distribution of light and its propagation direction; such a richness of information is used, during data processing, to reconstruct three-dimensional snapshots of the scene of interest. In traditional light-field imaging devices \cite{ng2005light,ng2006digital,georgiev2006light,georgiev2010focused}, based on the insertion of an array of micro-lenses before the sensor, directional resolution can be gained only at the expense of spatial resolution \cite{georgiev2006spatio}. Nonetheless, aided by additional post-processing algorithms \cite{dansereau2013decoding,perez2014super,li2016scalable}, they find applications in the most diverse fields, such as photography \cite{ng2005fourier} and microscopy \cite{broxton2013wave}. Plenoptic or light-field cameras based on intensity measurement can currently operate at the essentially same rates of a standard camera, which evidently outperforms any correlation imaging technique (whether volumetric or not) at similar resolutions. However, many interesting applications of light-field imaging in science, including the neuronal activity detection accomplished in Refs.~\cite{prevedel2014simultaneous,vaziri2018}, are performed at image rates ranging between $10$ and $100\,\mathrm{Hz}$ \cite{vaziri2018,ZHU2021130638}. The results presented in this paper thus demonstrate the competitiveness of correlated-photon imaging in these tasks: CPI based on correlated photons from chaotic sources brings a net advantage with respect to state-of-the-art light-field cameras, combining similar time performances with an improved volumetric resolution at analogous numerical aperture \cite{dilena2020correlation,scattarella2022resolution}. To the best of our knowledge, the presented CPI setup is indeed the first case of a feasible light-field device employing a single lens.

The article is organized as follows. In the ``Results'' section, we summarize the working principle of the novel CPI device and show the possibility to obtain light-field images with a $0.1\,\mathrm{s}$ acquisition time. In the ``Discussion'' section, we highlight the relevance of our results along with their implications for practical applications, and discuss both the current limitations of our experiment and the foreseen future developments. In the ``Materials and Methods'' section, we present details about the experimental setup. The Supplementary Information document further details on the measured correlation functions and the refocusing process, as well as complementary results on the interdependence between the number of collected frames (hence, the image acquisition time) and the SNR.

\section*{Results}

\begin{figure}
	\includegraphics[width=0.9\textwidth]{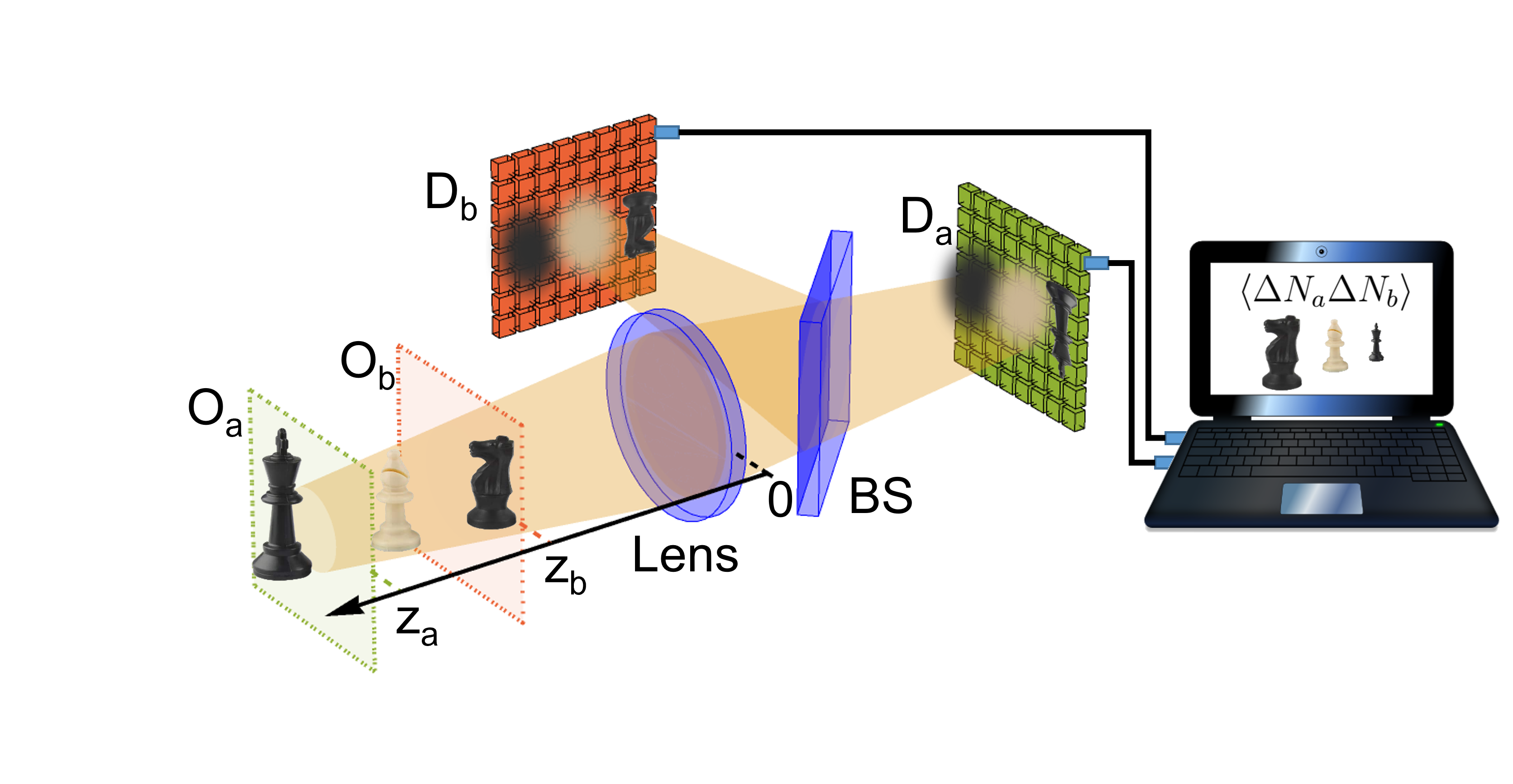}
	\caption{Working principle of the developed CPI protocol. $\mathrm{O}_a$ and $\mathrm{O}_b$ are the two conjugate planes of the high-resolution sensors $\mathrm{D}_a$ and $\mathrm{D}_b$; they are placed at the generic distances $z_a$ and $z_b$ from the lens, respectively. BS is a beam splitter sending light from the lens toward the two sensors. Pixel-by-pixel correlations between photon number fluctuations are evaluated by software and employed to reconstruct the volumetric image of the scene. See ``Materials and Methods'' for the detailed experimental setup.}
	\label{fig:setup}
\end{figure}

\begin{figure}
	(a)\includegraphics[width=.9\textwidth]{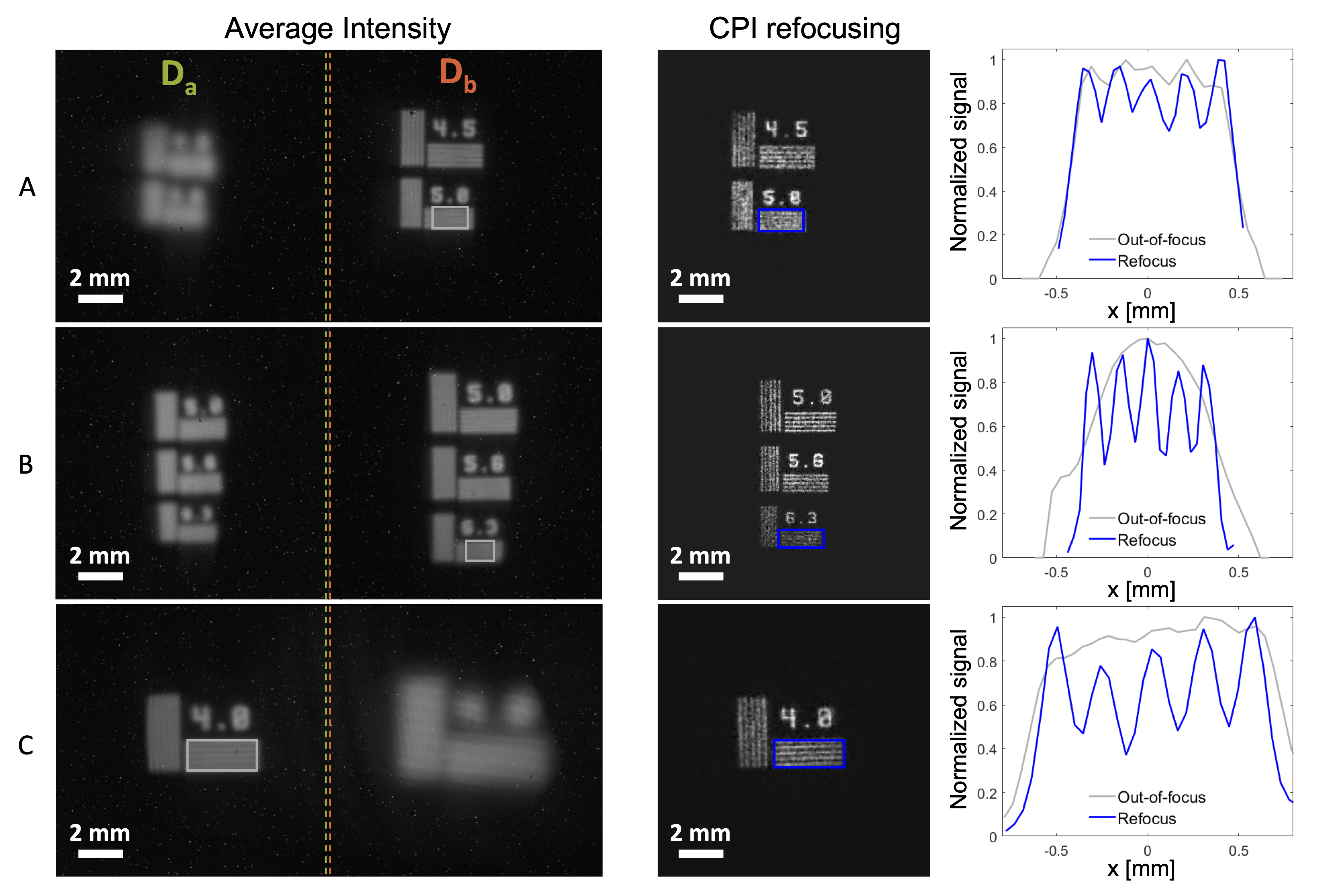}
	(b)\includegraphics[trim={0mm 0mm 0mm 10mm},clip,width=.9\textwidth]{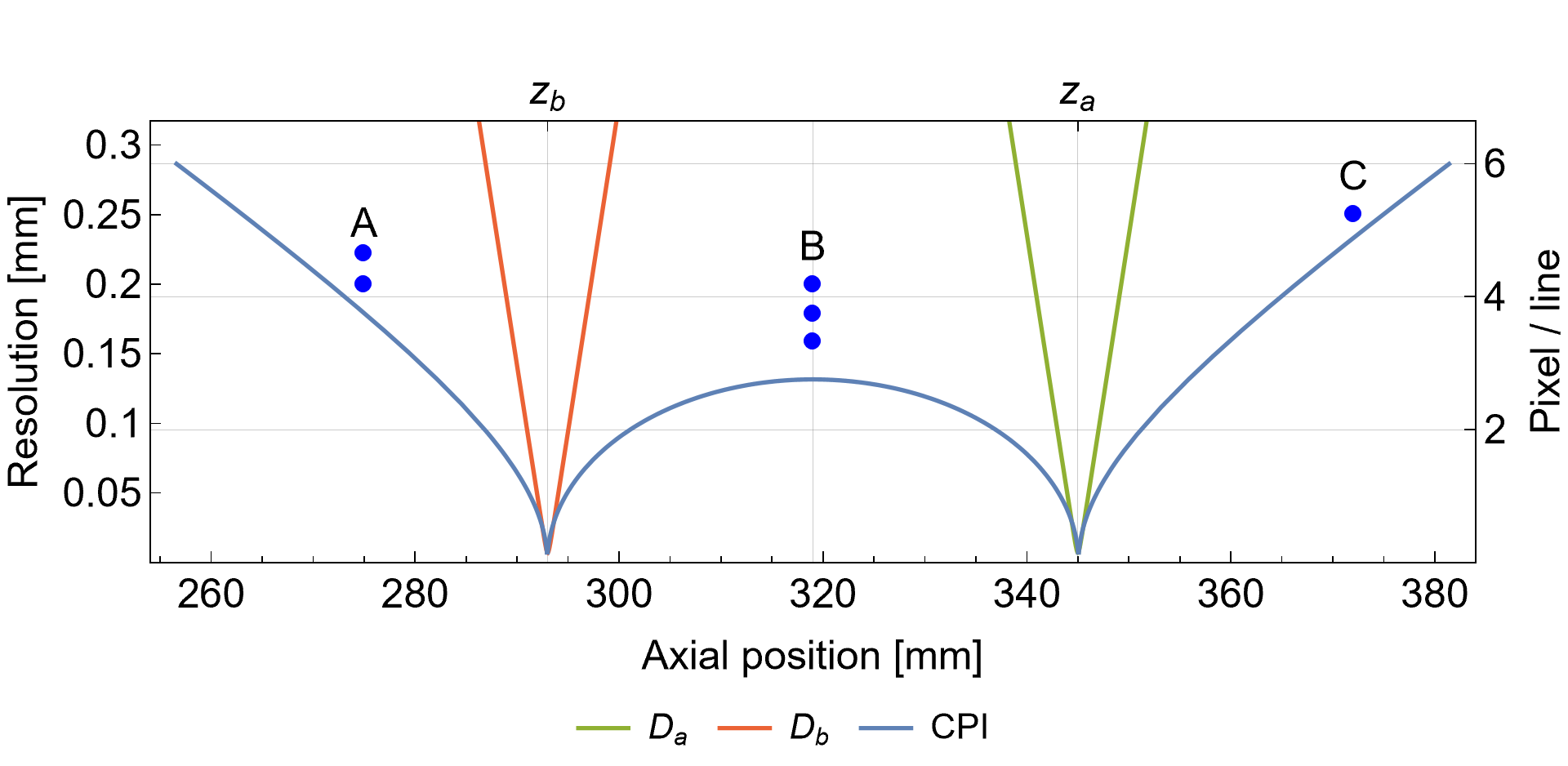}
	\caption{(a) Comparison between the images directly retrieved by the sensors, through their average intensity (left panels), and the images refocused by means of CPI (central panel); the plots in the right panel are obtained by integrating the highlighted rectangles in the corresponding average intensities (gray) and refocused images (blue) along the slit direction. All the reported data have been obtained by acquiring $N_t=9.8\times10^{3}$ frames, at full resolution, at $\sim9.8\times10^{4}$ frames per second, resulting in an overall acquisition speed of 10 volumetric images per second. 
	(b) Comparison of the resolution achieved by conventional imaging (orange for $\mathrm{D}_b$, green for $\mathrm{D}_a$) and CPI (blue) as a function of the axial distance from the lens. A ($z=275$ mm), B ($z=319$ mm) and C ($z=373$ mm) indicate the three different masks shown in panels (a); the plot shows their axial position and the distances between neighboring slits within the masks.}\label{fig:results}
\end{figure}

The working principle of the CPI camera is reported in Fig.~\ref{fig:setup} \cite{dilena2020correlation}. Two planes $\mathrm{O}_a$ and $\mathrm{O}_b$, chosen arbitrarily within the three-dimensional scene of interest, are focused on two high-resolution sensors, $\mathrm{D}_a$ and $\mathrm{D}_b$. Unlike a conventional light-field camera, involving both the usual camera lens and a micro-lens array, our CPI device is realized with a single lens, collecting light from both chosen planes, and focusing them on the two sensors.
Light from the scene is chaotic; hence, by computing the equal-time pixel-by-pixel correlation between the number of photons ($N_a$ and $N_b$) detected by the sensors $\mathrm{D}_a$ and $\mathrm{D}_b$, we obtain the correlation function:
    \begin{equation}\label{eq:correlation_def}
        \Gamma(\bm\rho_a,\bm\rho_b) = \langle N_a (\bm\rho_a) N_b (\bm\rho_b) \rangle - \langle N_a (\bm\rho_a) \rangle \langle N_b (\bm\rho_b) \rangle,
    \end{equation}
where $\langle\dots\rangle$ indicates the averaging process, while $\bm{\rho}_a$ and $\bm{\rho}_b$ are the coordinates identifying pixel positions on the sensors. The correlation function in Eq.~\eqref{eq:correlation_def} contains plenoptic information, and thus enables reconstructing 
features of a 3D object that can be placed both
between and beyond the two planes $\mathrm{O}_a$ and $\mathrm{O}_b$ imaged on the detectors \cite{dilena2018correlation, s22176665}. As explained in detail in the Supplementary Information, $\Gamma(\bm{\rho}_a,\bm{\rho}_b)$ encodes a collection of multi-perspective volumetric images; proper processing of these volumetric images provides the \textit{refocused image} of a specific transverse plane in the scene. Adopting chaotic light illumination entails that the magnitude of the correlation function \eqref{eq:correlation_def} scales like the product on the mean number of photons \cite{mandel1995optical}, thus being crucial in ensuring that correlation measurements provide analogous results both in the case of high-intensity, as in Ref.~\cite{massaro2022lightfield}, and in the single-photon regime, as in the present work. However, a key requirement in this sense, is that the SPAD array works in the linear regime (namely, the probability to detect a photon is proportional to the intensity of the impinging field), far from saturation.

Experimental results are reported in Fig.~\ref{fig:results}: Both sensors acquire blurred images of three different planar test targets (A, B, and C); the plenoptic information contained in the measured correlation function enables reconstructing the object details, in all three cases. In the panel on the left, we report the out-of-focus images of the test target, which is placed either within (case B) or outside (cases A and C) the volume defined by the two conjugate planes of the detectors; the effective refocusing enabled by CPI is shown in the center panels. The recovery in visibility deriving from refocusing is demonstrated in the right panels: here, we compare the linear images related with both average intensity and CPI, which are obtained by integration along the slit direction. CPI also enables over $10\times$ depth of field (DOF) enhancement at a resolution of 250 $\mu$m, and $12\times$ at 160 $\mu$m, with respect to a conventional imaging system with the same numerical aperture (NA). This can be seen by considering the curves reported in panel (b) of Fig.~\ref{fig:results}, together with the axial position and the distance between neighboring slits on the three test targets (A, B, and C) reported in panel (a). The plot shows the expected resolution limit of the refocused images (blue line), with varying axial position $z$, compared with the analogous limits associated with the conventional images focused on $\mathrm{D}_a$ (green line) and $\mathrm{D}_b$ (orange line). In particular, the blue line indicates the object detail size (i.e., the resolution) that can be refocused by our CPI device with 10\% visibility, as a function of the longitudinal distance ($z$) of the object from the lens. The green and red lines represent the natural DOF (as determined by the circle of confusion) of the images separately observed on the two sensors D$_a$ and D$_b$, at the given resolution.

The images reported in Fig.~\ref{fig:results} have been obtained at an overall acquisition speed of $10$ volumetric images per second. This is an unprecedented result in the field of correlation imaging, and indicates the feasibility of correlated-photon imaging at video rate. The noise analysis reported in the Supplementary Information shows the robustness of the developed technique: when the acquisition speed is reduced to 1 image per second by increasing the number of acquired frames by 1 order of magnitude, the SNR increases by nearly 35\%, and reaches its maximum value. A comparison between the CPI images acquired at the speed of 1 and 10 volumetric images per second is reported in Fig.~S3 of the Supplementary Information.

\section*{Discussion}

We have presented a quantum-inspired imaging system capable of collecting 10 plenoptic images per second. Since plenoptic images are a collection of multi-perspective volumetric images, they enable changing, in post-processing, the focusing plane within the entire axial range enclosed by the blue curve in Fig.~\ref{fig:results}b). This result entails a large reduction of the gap in time-performance between CPI and conventional light-field imaging, that is generally performed at a speed between 10 and 100 Hz, in scientific applications \cite{vaziri2018,ZHU2021130638}, but with a significant loss of resolution due to microlens array and intensity measurement. 

The key element to achieve such a critical improvement in the acquisition time is the integration of the SwissSPAD2 sensor in a chaotic-light based correlaton plenoptic imaging setup. This SPAD array enables to collect, with single-photon sensitivity, all the frames that contribute to the plenoptic correlation image, at a rate of almost $100.000$ frames per second. Such a fast rate is combined with both a resolution comparable to that of ordinary detectors and low noise (see Refs.~\cite{ulku2019spad,ulku2020spad,abbattista2021towards} for a detailed description of the sensor). The low noise of the detector is a key aspect for keeping as low as possible the number of frames $N_t$ required for reconstructing light statistics and correlations. Our SwissSPAD2 sensor has an on-board DDR3 memory bank (2 GB) that can be filled with a maximum of $131\,072$ measured binary frames. By saving the acquired data on the internal memory instead of streaming to an external disk, we were able to exploit the maximum speed of 97.7 kHz at full resolution. However, the limited capacity of the memory has bound us to single-image acquisitions instead of videos. In the future, we shall employ a new generation of SwissSPAD2 capable of streaming data from a $512 \times 512$ sensor to a workstation at full speed.

The new generation of SwissSPAD2 also provides a relevant improvement in the gating time, which can be reduced down to about 10 ns. This represents an extremely important parameter in correlation imaging based on chaotic light, since reconstruction of light statistics is optimal when the detector exposure time matches the coherence time of light \cite{mandel1995optical}. The possibility to match coherence times as small as hundreds of ps would open the way to CPI with broadband sources, thus leading the way toward passive quantum imaging devices.

It is reasonable to expect the achieved acquisition speed to be further increased through computational techniques enabling to use less frames to achieve a comparable SNR; examples are compressive sensing \cite{katz2009compressive,jiying2010high}, quantum tomography \cite{vrehavcek2002maximum}, and machine learning \cite{Li:21}. All these techniques are currently being developed in the framework of CPI, and we plan to integrate them with our refocusing algorithm~\cite{abbattista2021towards}. To further increase the SNR while reducing the acquired number of frames, we are also working toward employing, within the data analysis, the statistical properties of the correlation function, in a similar fashion as in ~\cite{ferri2010differential}. It is interesting to emphasize, however, that all data presented in this work have not been treated with any denoising algorithm, or post-processing method, other than the refocusing algorithm described in the Supplementary Information.

The novelty of the implemented CPI setup also stands in the fact that two arbitrary planes are focused on the two sensors~\cite{dilena2020correlation}, as opposed to conventional approaches involving imaging of the main lens for retrieving directional information. This approach enables: (i) parallel acquisition of two diffraction limited images within the three-dimensional scene of interest, (ii) \textit{single-lens light-field imaging}, which is quite significant considering the disadvantages and physical limitations connected with the use of micro-lenses (i.e., resolution loss and reduced 3D imaging capability), (iii) a DOF enhancement by over 1 order of magnitude, without sacrificing diffraction-limited resolution.

\section*{Materials and Methods} 

\begin{figure}
	\includegraphics[width=0.9\textwidth]{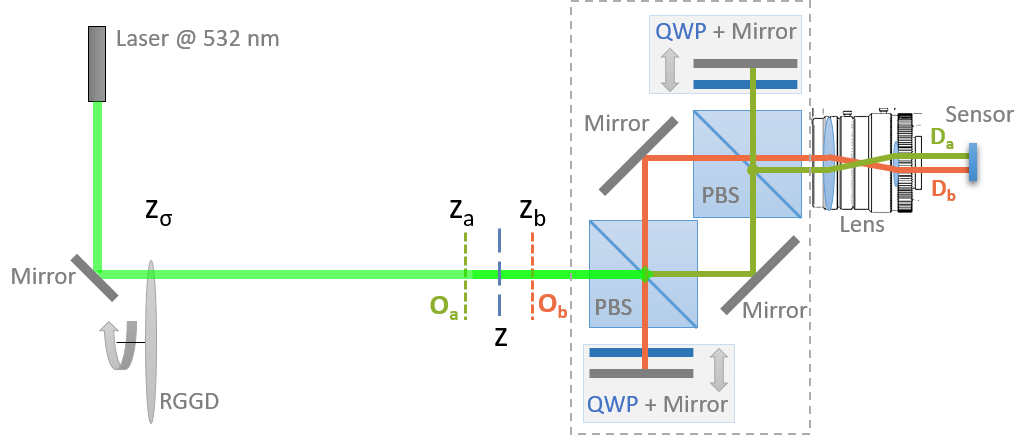}
	\caption{Technical scheme of the developed CPI setup. The chaotic source is made of a diode laser illuminating a rotating ground glass disk. Two planes, $\mathrm{O}_a$ and $\mathrm{O}_b$, arbitrarily chosen in the surrounding of the scene of interest, are imaged by a unique lens onto two disjoint high-resolution detectors $\mathrm{D}_a$ and $\mathrm{D}_b$, which are practically implemented by using two halves of the same SwissSPAD2 sensor. Two optical paths, one for each detector, are realized by means of two polarizing beams splitters (PBS), two quarter-wave plates (QWP) and four mirrors; each pair of QWP and mirror is mounted on a translation stage, which offers flexibility in the choice of the two planes $\mathrm{O}_a$ and $\mathrm{O}_b$, when preparing the acquisition.}\label{fig:details}
\end{figure}

The optical system is aimed at maximizing speed of acquisition and performance in terms of resolution versus DOF trade-off, while guaranteeing flexibility in the focusing capability. The design of the employed CPI setup is oriented to the acquisition of generally demagnified images, as in an ordinary camera. The lack of two synchronized SPAD arrays has imposed using two halves of a single SPAD array as the two sensors $\mathrm{D}_a$ and $\mathrm{D}_b$. This entails some constraints to the setup design: demagnified images can only be obtained if the two CPI paths are separated upstream of the lens, rather than downstream (as reported in Fig.~\ref{fig:setup}, and originally proposed in Ref.~\cite{dilena2018correlation}). The experimental setup thus consists of two main parts (Fig.~\ref{fig:details}):
\begin{enumerate}
	\item the ultra-fast imaging device, made of a camera lens (Navitar MVL75M1, of focal length $75\,\mathrm{mm}$ and focal ratio $2.8$) mounted on the SPAD array sensor;
	\item a ``CPI adapter'', represented in Fig.~\ref{fig:details} by the dashed gray rectangle.
\end{enumerate}

The CPI adapter endeavors the ultra-fast camera with plenoptic properties by first creating (through the first polarizing beam-splitter) and then recombining (through the second polarizing beam-splitter) two optical paths, which we shall indicate as $a$ (depicted in green) and $b$ (depicted in red). Each optical path contains a delay line, offering the required flexibility for choosing the two arbitrary planes $\mathrm{O}_a$ and $\mathrm{O}_b$, when preparing the acquisition. In our setup the distances between the planes $\mathrm{O}_{a,b}$ and the lens are $z_a=345$ mm and $z_b=293$ mm, respectively.
The delay lines are made by the combined system (QM$_{a,b}$) of a quarter-wave plate (QWP) and a mirror. This combined system converts light from H-polarized to V-polarized, and viceversa, so that the beam that is back-reflected by QM$_{a,b}$ is then reflected/transmitted by the corresponding PBS toward the camera lens. Changing the optical path in arms $a$ and $b$ defines the specific plane to be imaged on sensor $\mathrm{D}_a$ and $\mathrm{D}_b$, respectively. In fact, given the lens focal length $f$ and the fixed lens-to-sensor distance $z_i$, the distance $z_o$ of the object plane from the lens is uniquely defined by the thin lens equation.
Hence, the two planes $\mathrm{O}_a$ and $\mathrm{O}_b$, imaged on $\mathrm{D}_a$ and $\mathrm{D}_b$, respectively, are both placed at an optical distance $z_o$ from the lens; however, the actual planes that are imaged on two disjoint halves of the sensor, are determined by length of the delay lines, which enable to arbitrarily choose the distances $z_a$ and $z_b$, associated with two different planes within the volume of interest.
We should specify that the versatility in choosing the two planes is a useful feature when setting up the acquisition, since it allows the experimenter both to select the planes to be focused and to define the volume that can be refocused (as defined by the blue curve in Fig.~\ref{fig:results}(b)); the specific choice, however, plays no role during the acquisition itself.
Both delay lines are characterized by the same magnifications $M=-z_i/z_o$, numerical aperture (NA), and resolution at focus, as defined by the camera lens. The clear aperture of both the polarizing beam splitter, PBS (45 mm), and the optics (2 inches) in the delay lines have been chosen to enable fully exploiting the NA of the camera lens. 
In order to maximize its fill factor, the SwissSPAD2 sensor is equipped with a microlens array; its NA $\approx 0.25$ is larger than the lens NA on the image side ($\text{NA}_i=\text{NA}_o/|M| = 0.13$) and does not limit the NA of the CPI device. However, the pixel size of the sensor is larger than the achievable diffraction limited resolution; hence, the setup has a pixel limited resolution of 95 $\mu$m.

In the present experiment, the CPI device was employed to image transmissive planar test targets placed out of focus, as shown in Fig.~\ref{fig:results}. The targets are illuminated by a chaotic light source of controllable polarization, intensity, and coherence time, made by a green diode laser (Thorlabs CPS532, $\lambda=532$ nm) scattered by a rotating ground glass disk (GGD). At the maximum rotation speed of the GGD (30 Hz), the measured coherence time of the source is $t_\text{ch}\simeq 15 \,\mu\mathrm{s}$.

SwissSPAD2 employs a design that provides one of the largest resolutions 
($512 \times 512$ photodiodes operating in Geiger mode) as well as one of the highest sensitivity (50\% photon detection probability at 520 nm) and lowest dark count rate ($0.26\text{ cps}/\mu\text{m}^2$, equivalent to a median value of less than 10 cps per pixel) combinations among SPADs which are built with standard CMOS-process technologies. Its 10.5\% native fill factor is improved by 4-5 times, for collimated light, by means of the use of a microlens array. The output of each frame consists of a binary matrix identifying the pixels that have been triggered by at least one photon. Due to the binary nature of the signal, it is of utmost importance for the reconstruction of intensity correlations to work close to the linear regime \cite{antolovic2016photon,antolovic2018photon}, in which the probability to detect a photon is proportional to the intensity of the impinging electromagnetic field.

\section*{Supplementary information}

\renewcommand{\thefigure}{S\arabic{figure}}
\renewcommand{\thetable}{S\arabic{table}}
\renewcommand{\theequation}{S\arabic{equation}}
\renewcommand{\thesubsection}{S\arabic{subsection}}

\subsection{Plenoptic properties of the intensity correlation function}

As demonstrated in Ref.~\cite{dilena2020correlation}, plenoptic information is encoded in the correlation function of Eq. (1) in the main text, representing the correlation between the intensity fluctuations reaching two points, of which one is placed on the detector $\mathrm{D}_a$, and the other on the detector $\mathrm{D}_b$. The correlation function reads, up to irrelevant factors that do not depend on either $\bm{\rho}_a$ or $\bm{\rho}_b$,
\begin{equation}
\Gamma(\bm\rho_a,\bm\rho_b) =
\left|\iint A(\bm\rho_o)A^*(\bm\rho_o^\prime)
\Psi(\bm\rho_o, \bm\rho_o^\prime, \bm\rho_a, \bm\rho_b)d^2\bm\rho_o d^2\bm\rho_o^\prime\right|^2,
\label{eq:gamma}
\end{equation}
where $A$ is the aperture function of the object. The function $\Psi$ can be considered as a ``second-order point-spread function'', which determines the correspondence between object points and detector points. By referring to the parameters in Fig.~3 of the main text, and assuming the source emits chaotic light with an average intensity profile $S(\bm{\rho}_s)$ and negligible transverse coherence, we have
\begin{equation}
\Psi(\bm\rho_o, \bm\rho_o^\prime, \bm\rho_a, \bm\rho_b) =p_a (\bm\rho_o,\bm\rho_a)
p_b (\bm\rho_o^\prime,\bm\rho_b)
\int S(\bm\rho_s) e^{\frac{ik}{2z_\sigma}\left[\bm\rho_o^2 - (\bm\rho_o^\prime)^2 - 2(\bm\rho_o-\bm\rho_o\prime)\cdot \bm\rho_s \right]}d^2\bm\rho_s, 
\end{equation}
with the functions $p_j$, $j=a,b$, describing field propagation from the object to the detector. Calling $P$ the pupil function of the lens, we obtain
\begin{equation}\label{eq:pj}
p_j (\bm\rho_o,\bm\rho_j) =
\int P(\bm\rho_{\ell j})e^{i k_z \phi_j(\bm\rho_o,\bm\rho_{\ell j},\bm\rho_j)} d^2\bm\rho_{\ell j}
\end{equation}
with
\begin{equation}\label{eq:phi}
\phi_j (\bm\rho_o,\bm\rho_{\ell j},\bm\rho_j) =
\left(\frac{1}{z-z_j+z_o}-\frac{1}{z_o}\right)\frac{\bm\rho_{\ell j}^2}{2}-
\left(\frac{\bm\rho_o}{z-z_j+z_o}-\frac{\bm\rho_j}{M\,z_o}\right)\cdot\bm\rho_{\ell j} \quad.
\end{equation}
Notice that the differences in phases $\phi_j$ with respect to the original results described in Ref.~\cite{dilena2020correlation} are due to the fact that the optical paths are split upstream of the lens, and not downstream. 

The plenoptic properties of the correlation function are easily deduced, for example, from the fact that, by applying a stationary-phase approximation to the integrals which defines it, one obtains
\begin{equation}
	\Gamma(\bm\rho_a,\bm\rho_b) \sim 
	\left|A\left[
	\frac{(z_b-z)\bm\rho_a - (z_a-z)\bm\rho_b}{M\Delta z}
	\right]\right|^4 .
\end{equation}
Therefore, $\Gamma(\bm\rho_a,\bm\rho_b)$ represents a collection of images of the object, whose shifts and scaling depend on the axial position of the latter. The refocusing process, consisting in properly realigning the images contained in the correlation function, is discussed in the next section.

\subsection{Refocusing and correlation aperture}

\begin{figure}
	\centering
	\includegraphics[width=0.5\linewidth]{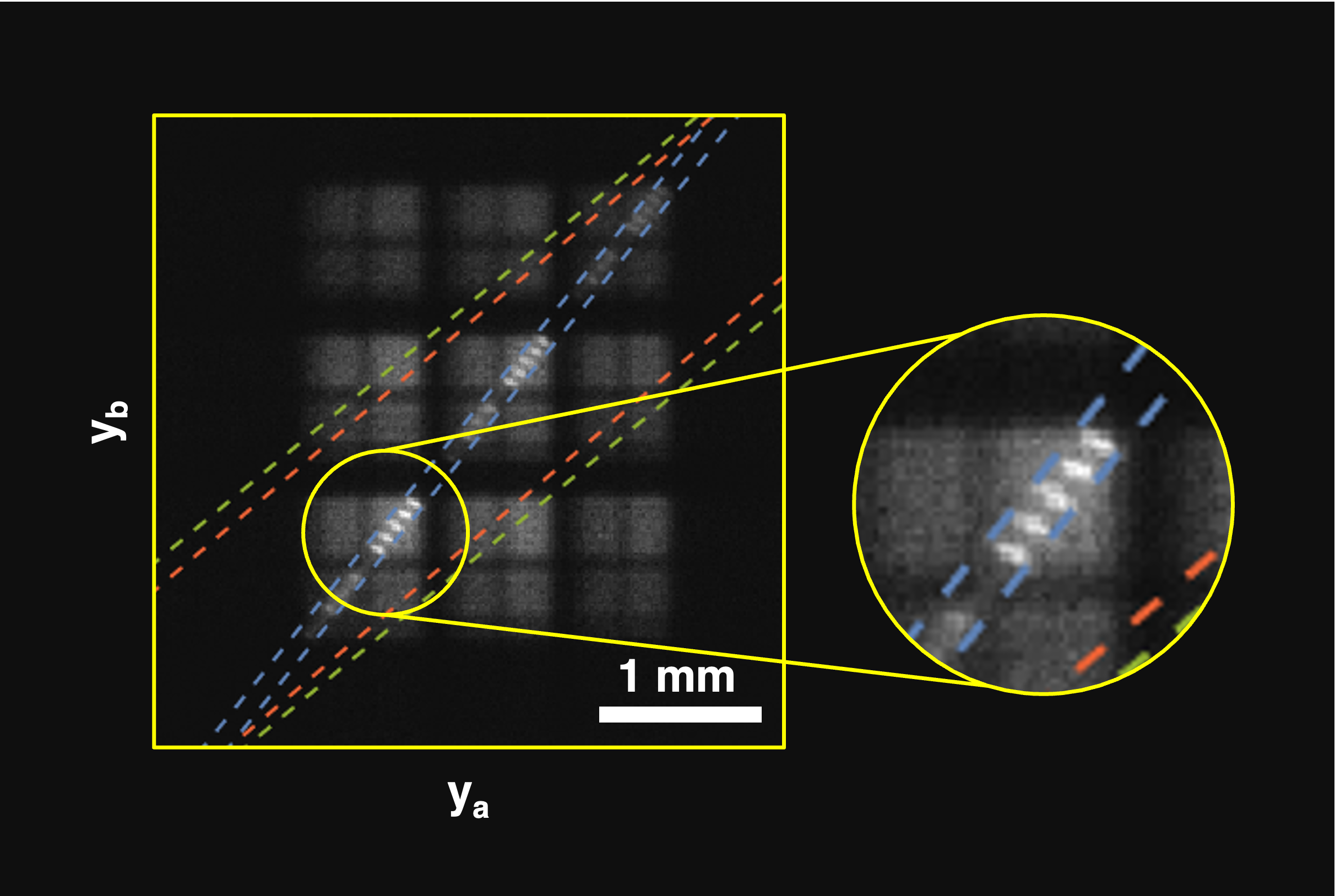}
	\caption{Two-dimensional correlation function obtained after integration along the $x$ direction of data acquired by the two sensors in the experiment corresponding to case B (as reported in Fig.~2 of the main text). Dashed green and blue lines identify the region (defined by the lens) where non-zero correlations can be found. The dotted blue lines identify the correlated region defined by the light source profile. The overall correlation region is given by the intersection of the three regions.}
	\label{fig:gamma}
\end{figure}

In a ray-optics approximation, the correlation function that is measured reads \cite{dilena2020correlation,pepe2016plenoptic,massaro2022effect}:

\begin{align}
	\Gamma(\bm\rho_a,\bm\rho_b) = &
	\left|A\left[
	\frac{(z_b-z)\bm\rho_a - (z_a-z)\bm\rho_b}{M\Delta z}
	\right]\right|^4
	\nonumber \\ &\times 
	\left|P\left[
	\frac{(z_o+\Delta z)\bm\rho_a - z_o\bm\rho_b}{M\Delta z}
	\right]\right|^2
	\left|P\left[
	\frac{z_o\bm\rho_a - (z_o-\Delta z)\bm\rho_b}{M\Delta z}
	\right]\right|^2
	\nonumber \\ &\times
	\left|S\left[
	\frac{(z_b-z_\sigma)\bm\rho_a - (z_a-z_\sigma)\bm\rho_b}{M\Delta z}
	\right]\right|^2
	\quad.
\label{eq:gamma:ray}
\end{align}
This equation shows that $\Gamma$ contains information regarding the object aperture $A$, and that also the lens aperture $P$ and source intensity profile $S$ have effects on the correlation function. Each aperture defines a region inside the correlation space $(\bm\rho_a,\bm\rho_b)$, as shown in Fig.~\ref{fig:gamma}: the green and red lines indicate the regions where non-zero correlations can be expected, as defined by the lens; the blue lines, instead, indicate the effects of limited aperture of the source, \textit{i.e.} the width of the Gaussian profile illuminating the GGD. In fact, correlation can only be measured if the light passes through all the apertures, and this means that the correlation region is the intersection of the three correlation regions defined by $P$ and $S$. 
By considering only the correlation region, in data analysis, one eliminates all the spurious signal outside that area, which can only be due to noise. Moreover, to simplify the numerical operations, one can consider only the smallest of the regions defined by $P$ and $S$ rather than their intersection (in our case, the area defined by the source). In order to determine the dominating aperture, an effective radius in the correlation space should be evaluated for all the apertures. For the lens of aperture $P$ we consider a circular 2D pupil function with radius equal to the effective radius of the lens: $r_\ell=\text{NA}_o z_o$.
If we call \emph{correlation aperture} (CA) the radius of the region defined by the lens, we obtain
\begin{equation}\label{eq:ca}
	\text{CA}_{\text{lens},a} =
	\text{NA}_o \frac{|M\Delta z|}{\sqrt{\left(1 + \Delta z/z_o\right)^2+1}}
	\quad \text{CA}_{\text{lens},b} =
	\text{NA}_o \frac{|M\Delta z|}{\sqrt{\left(1 - \Delta z/z_o\right)^2+1}} .
\end{equation}
The same quantity can be defined and calculated for the source:
\begin{equation}
\text{CA}_\text{source} =
\frac{r_\sigma}{z_\sigma}\frac{|M\Delta z|}{\sqrt{(z_b/z_\sigma-1)^2+(z_a/z_\sigma-1)^2}}
\quad,
\label{eq:ca:source}
\end{equation}
where $r_\sigma=c\,\sigma=1.44\text{ mm}$ is the radius of the source aperture, that depends on the standard deviation $\sigma=1.02\text{ mm}$ of its Gaussian profile. 
The factor $c$ is used as an optimization parameter to maximize the SNR. To show that the source is the limiting aperture in our experimental conditions, we report the three radii:
\begin{equation}
\text{CA}_{\text{lens},a} = 0.63\,\mathrm{mm}, \quad
\text{CA}_{\text{lens},b} = 0.53\,\mathrm{mm}, \quad
\text{CA}_\text{source}   = 0.07\,\mathrm{mm}.  
\label{eq:ca:num}
\end{equation}

The refocusing $\alpha(z)$ is a linear operator that transforms the coordinates on the detector planes $\bm\rho_a$ and $\bm\rho_b$ into two new ones: $\bm\rho_r$ related to the refocusing plane, and $\bm\rho_s$ related to another generic plane in $z_s$. It reads:
\begin{equation}
\alpha(z):
\begin{bmatrix}
    \bm\rho_a \\
    \bm\rho_b
\end{bmatrix}
\mapsto
\begin{bmatrix}
    \bm\rho_r \\
    \bm\rho_s
\end{bmatrix}
=\frac{1}{M\Delta z}
\begin{bmatrix}
    z_b-z & z-z_a\\
    z_b-z_s & z_s-z_a
\end{bmatrix}
\begin{bmatrix}
    \bm\rho_a \\
    \bm\rho_b
\end{bmatrix}
\label{eq:refocus}
\quad.
\end{equation}
By applying the transformation to the measured correlation function in the position $z$ of the target, one obtains
\begin{equation}\label{eq:gamma:ref}
	\Gamma_r(\bm\rho_r,\bm\rho_s,z) =  \Gamma\left(
	{\mbox{\large $\alpha^{-1}(z)$}}
	{\footnotesize
	\begin{bmatrix}
		\bm\rho_r \\
		\bm\rho_s
	\end{bmatrix}
	}\right) \sim 
	\left|A\left(\bm\rho_r\right)\right|^4
	\left|S\left(\bm\rho_s\right)\right|^2
	\quad.
\end{equation}
The last approximate equality stems from the much smaller radius of the source with respect to the lens, so that the latter can be considered irrelevant. We should point out, however, that the algorithm of Eq.\eqref{eq:refocus} is the most convenient choice only when the source is the limiting aperture and should be modified accordingly if the lens becomes the dominating aperture. The final integration
\begin{equation}
	\Sigma(\bm\rho_r,z)=\int\Gamma_r(\bm\rho_r,\bm\rho_s,z) d^2\bm{\rho}_s \sim 
	\left|A\left(\bm\rho_r\right)\right|^4
\label{eq:sigma:ref}
\end{equation}
can be done by limiting the integration domain to an area defined by the source radius.

By comparing  Eqs.~\eqref{eq:gamma:ray} and \eqref{eq:refocus}, we see that the coefficients on which the object aperture depends correspond to the first row of the refocusing matrix, while the second row contains the coefficients that appear in the source profile. The first line gives us an insight into what the refocusing algorithm actually does, that is, rescaling the detector coordinates so that, in the transformed plane, the object depends only on the set of two-dimensional coordinates $\bm\rho_r$. It is also clear that the second line, defining the variable $\bm\rho_s$ that is integrated in Eq.~\eqref{eq:sigma:ref}, plays no relevant role for the object reconstruction, and can be chosen arbitrarily. However, some choices are more convenient than other when defining the integration variable. For example, since the extension of the correlation function is defined by the size of the apertures, a clever choice is to define $\bm\rho_s$ as the transverse coordinate of one of the limiting apertures. By doing so, one can integrate only where non-zero correlations are expected, and limit the integration of Eq.~\eqref{eq:sigma:ref} to a much smaller area than what has been measured, so as to speed up calculations quite dramatically. Given the values of the correlation apertures in Eq.~\eqref{eq:ca:num}, it made sense to us to choose the integration variable as the one defined by the source. If that were not the case, the second line would have been modified with the coefficients of the aperture defining the major limitation on the correlation area.

\subsection{Data analysis workflow}

Data analysis has been performed on MATLAB and consisted of two main parts, that are data reading and correlation, and refocusing. In the first part, the binary frames are read from the disk and the correlation function is evaluated by averaging over all frames. Computation of the correlation function is fast and efficient, so that the speed of the process was mostly defined by the data read rate of the disk; reading $\sim 10^4$ frames ($512 \times 256$) and computing the correlation function took about $200$ s. After the correlation function is available in the workstation memory as a 4D array, the operation of refocusing took about $14$ s per axial coordinate $z$.

\subsection{Study of the SNR dependence on the number of frames}

\begin{figure}
    \centering
    \includegraphics[width=0.7\textwidth]{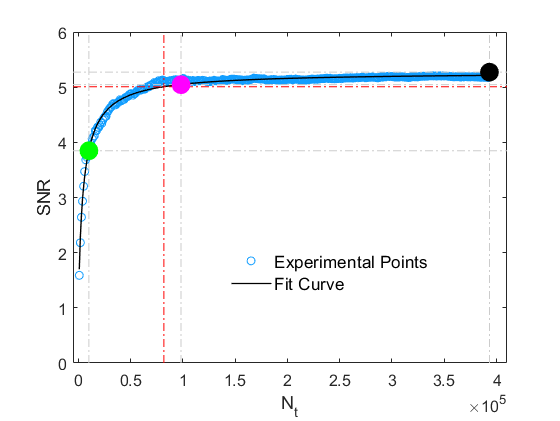}
    \caption{Behavior of the SNR in the image refocused by CPI (referred to case C shown in Fig.~3 of the main text), as a function of the number of collected frames $N_t$. Blue circles represent the SNR evaluated on the experimental refocused images, while the solid black line corresponds to the fitting curve $(a+b/N_t)^{-1/2}$. The points discussed in the text, and corresponding to $N_t=4.0 \times 10^5$, $N_t=9.8\times 10^4$, and $N_t=9.8\times 10^3$, are highlighted in black, magenta, and green, respectively. The red dotted lines correspond to $N_t=8 \times 10^4$, where we estimated a decrease of the SNR by 5\% of its maximum.}
    \label{fig:snr_plot}
\end{figure}

\begin{figure}
\centering
\subfigure[$N_t=9.8\times10^3$, SNR $=3.9$]{\includegraphics[width=0.33\textwidth]{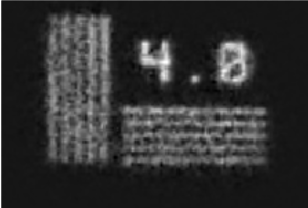}}
\subfigure[$N_t=9.8\times10^4$, SNR $=5.1$]{\includegraphics[width=0.33\textwidth]{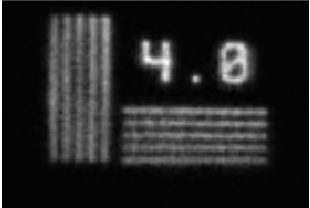}}
\subfigure[$N_t=4.0\times10^5$, SNR $=5.3$]{\includegraphics[width=0.33\textwidth]{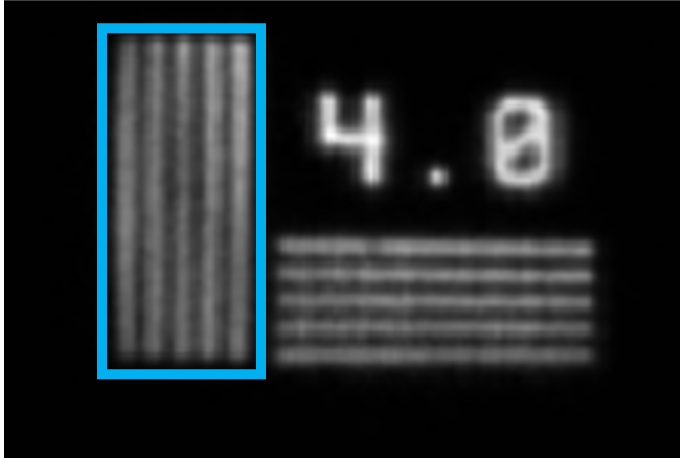}} 
\caption{Comparison of the refocused images acquired at 10 (a) and 1 (b) volumetric images per second. Panel (c) shows the refocused image using the whole acquired dataset. The cyan rectangle indicates the area where the SNR has been evaluated on the three images.}
\label{fig:speed}
\end{figure}

We report in Fig.~\ref{fig:snr_plot} the full analysis of the dependence of the SNR on the number of frames $N_t$. The analysis has been performed in the experimental conditions corresponding to case ``C'' of Fig.~2 (main text). The experimental points indicates that the SNR tends to saturate as the number of frames increases. This behavior deviates from the $\sqrt{N_t}$ scaling, expected if the collected frames were statistically independent, since our source can only provide a finite number of samples. We thus fit the experimental points of the SNR with the function
\begin{equation}\label{eq:snr}
\text{SNR}(N_t) = \frac{1}{\sqrt{a+\frac{b}{N_t}}} ,
\end{equation}
which accounts for both the $\sqrt{N_t}$ behavior at a small number of frames and the saturation at a high number of frames. The best fit provides $a=3.60\times 10^{-2}$ and $b=3.18\times 10^2$.
The plot in Fig.~\ref{fig:snr_plot} indicates that variations in the number of acquired frames around $N_t=4.0 \times 10^5$ does not yield an appreciable improvement in image quality. Actually, a reduction of $N_t$ by a factor close to $4$ ($N_t=9.8\times10^4$) with respect to the whole dataset, which enables to collect each CPI image in one second ($T_{\mathrm{CPI}} = 1\,\mathrm{s}$), produces a decrease in the SNR by less than $5\%$. Moreover, a further reduction in the number of frames to $N_t=9.8\times10^3$, which enables to collect 10 CPI in one second ($T_{\mathrm{CPI}} = 0.1\,\mathrm{s}$), Fig.~\ref{fig:speed} shows a comparison between these three cases, with the corresponding estimated values of the SNR.
The SNR in the region of interest enclosed in the cyan rectangle highlighted in Fig.~\ref{fig:speed}, has been estimated as 
\begin{equation}
    \text{SNR} = \frac{ \overline{\Sigma}_{\mathrm{in}} }{ \Delta_{\mathrm{in}} \Sigma } ,
\end{equation}
where $\overline{\Sigma}_{\mathrm{in}}$ represents the average value of the signal, in the image, in correspondence of the transmissive parts of the considered five-slit group; the denominator $\Delta_{\mathrm{in}} \Sigma$ represents the standard deviation of the values contributing to the numerator. Such a definition relies on the realistic assumption that the statistical distributions of the signal in the refocused image, in correspondence of the transmissive parts, are identical. 

\section*{Author contributions statement}

Conceptualization: EC and CB (sensor), MD. Methodology: EC and CB (sensor), MD, FD, GM, FVP, SV. Software: AU and PM (sensor), FD, FS, GM. Firmware: AU and PM. Theoretical analysis and modeling: FD, DG, GM, FVP. Validation: PM (sensor), SV. Setup implementation: PM (sensor), SV. Data acquisition: SV. Data analysis: FD, GM, SV, FS. Interpretation of the results: FD, MD, GM, FVP. Supervision: EC, CB, MD, FVP. Funding acquisition and project management: CB and MD. Writing—original draft preparation: FD, FS, FVP, MD, CB. writing—review and editing: all authors.

\end{document}